\documentclass[conference]{IEEEtran}
\IEEEoverridecommandlockouts

\usepackage{color}
\usepackage{graphicx}
\usepackage{epstopdf}
\usepackage{amsmath}
\usepackage{amssymb}
\usepackage{algorithm}
\usepackage{algorithmic}
\usepackage{multirow}
\usepackage{booktabs}
\usepackage{array}
\usepackage{amsthm}
\usepackage{stfloats}
\usepackage{caption}
\usepackage{subfigure}
\usepackage{bm}
\usepackage{setspace}
\usepackage{diagbox}
\usepackage{flushend}

\newcommand{\be}{\begin{equation}}
\newcommand{\ee}{\end{equation}}
\newcommand{\bea}{\begin{eqnarray}}
\newcommand{\eea}{\end{eqnarray}}
\newcommand{\ba}{\begin{array}}
\newcommand{\ea}{\end{array}}

\allowdisplaybreaks[4]

\newcounter{TempEqnCnt}

\def\BibTeX{{\rm B\kern-.05em{\sc i\kern-.025em b}\kern-.08em
    T\kern-.1667em\lower.7ex\hbox{E}\kern-.125emX}}
\begin{document}

\title{Joint EM-BB Beamforming Design for ISAC Systems With Reconfigurable Antenna Arrays}

\author{ \IEEEauthorblockN{Jvxiang Hu$^{\dag}$, Mengzhen Liu$^{\dag}$, Rang Liu$^{\ddag}$, Qian Liu$^{\dag}$, and Ming Li$^{\dag}$ \\
		\vspace{-0.3 cm} }
	\IEEEauthorblockA{$^{\dag}$ Dalian University of Technology, Dalian, Liaoning 116024, China \\
		E-mail: \texttt{\{hujvxiang, liumengzhen\}@mail.dlut.edu.cn, \{mli, qianliu\}@dlut.edu.cn}}
	
	\IEEEauthorblockA{$^{\ddag}$Friedrich-Alexander-University Erlangen-Nuremberg (FAU), Erlangen 91058, Germany, 
	E-mail: \texttt{rang.liu@fau.de}}

    \vspace{-0.5cm}	
        
        }

\maketitle
\pagestyle{empty}  
\thispagestyle{empty} 

\begin{abstract}
This paper investigates the joint electromagnetic (EM)-domain and baseband (BB)-domain beamforming for integrated sensing and communication (ISAC) systems enabled by pattern- and polarization-reconfigurable pixel-antenna arrays. A multiport-network-based antenna model is developed to map each binary switch configuration to its radiation response using the impedance and open-circuit radiation data extracted from Computer Simulation Technology (CST) Studio Suite simulations. Based on this model, polarization-aware virtual channels are constructed for downlink communication and statistical target sensing, leading to a mixed-integer nonconvex problem that maximizes the average sensing signal-to-noise ratio (SNR) subject to user signal-to-interference-plus-noise ratio (SINR) and transmit-power constraints. To handle the coupled binary and continuous variables, we propose an alternating optimization algorithm, in which the EM-domain switch-state matrices are updated using successive exhaustive Boolean optimization (SEBO), the BB-domain precoder is optimized via a minorization-maximization (MM) and second-order cone programming (SOCP) procedure, and the sensing combiner is obtained from a Rayleigh-quotient update. Simulation results show that the proposed design achieves higher sensing SNR and stronger target-angle robustness than the fixed-patch benchmark.
\end{abstract}

\vspace{0.08cm}

\section{Introduction}
 
Integrated sensing and communication (ISAC) enables a common wireless transceiver to reuse spectrum, hardware, and waveform resources for both data transmission and environmental sensing. Existing ISAC designs mainly rely on baseband (BB) beamforming to trade off target illumination and user quality-of-service (QoS). However, once the antenna hardware is fixed, its electromagnetic (EM) characteristics, including radiation pattern and polarization responses, impose intrinsic limits on the available spatial and polarization degrees of freedom (DoFs). To overcome this limitation, joint EM-BB beamforming leverages reconfigurable antennas to reshape the EM fields at the aperture before BB signal processing, opening a new design dimension for ISAC systems.

Reconfigurable antennas reshape EM responses through tunable hardware and provide additional DoFs beyond BB beamforming \cite{Costantine, Mengzhen CM}. Recent tri-hybrid designs further integrate EM-domain reconfigurability with RF/BB beamforming \cite{Castellanos2026TriHybrid,liu2025tritimescale}. Among various implementations, pixel antennas exploit switch-controlled topology changes to enable programmable polarization and pattern response \cite{MurchPixel}. Such reconfigurability further supports antenna coding \cite{Shen2025AntennaCodingPixel}, including multi-user and spatial-multiplexing applications \cite{hongyu2027, Han2026PixelSpatialMultiplexing}.

Despite recent progress, most EM-domain beamforming studies remain communication-oriented, targeting channel gain, radiation-pattern shaping, or spectral-efficiency maximization. These objectives do not fully capture the requirements of ISAC, where the transmit array must illuminate sensing targets while maintaining user signal-to-interference-plus-noise ratio (SINR), and the receive array must enhance weak echoes before BB combining. Moreover, radiation pattern and polarization provide complementary EM-domain DoFs: the former controls angular selectivity, whereas the latter is important for maintaining sensing capability under polarization mismatch \cite{Ma2025PolarizedISAC}. Therefore, an ISAC-oriented joint EM-BB framework is needed to jointly optimize transmit/receive switch configurations and BB beamformers under communication-SINR and sensing signal-to-noise ratio (SNR) metrics.

In this paper, we study joint EM-BB beamforming for ISAC systems equipped with pattern- and polarization-reconfigurable pixel-antenna arrays. Using microwave multiport-network theory and antenna data extracted from Computer Simulation Technology (CST) Studio Suite, we map binary switch configurations to polarization-resolved radiation responses and construct polarization-aware virtual channels for communication and statistical sensing. We formulate an average sensing-SNR maximization problem under user-SINR and transmit-power constraints, and develop an alternating algorithm based on successive exhaustive Boolean optimization (SEBO), minorization-maximization (MM), and second-order cone programming (SOCP) to jointly update the transmit/receive switch configurations, BB precoder, and sensing combiner. CST-based simulations show that the proposed design improves sensing SNR and target-angle robustness over fixed-radiation benchmark.

\section{System Model and Problem Formulation}
\subsection{Single Pixel Antenna Modeling}
\begin{figure}[!t]
  \centering
  \includegraphics[width=0.9\linewidth]{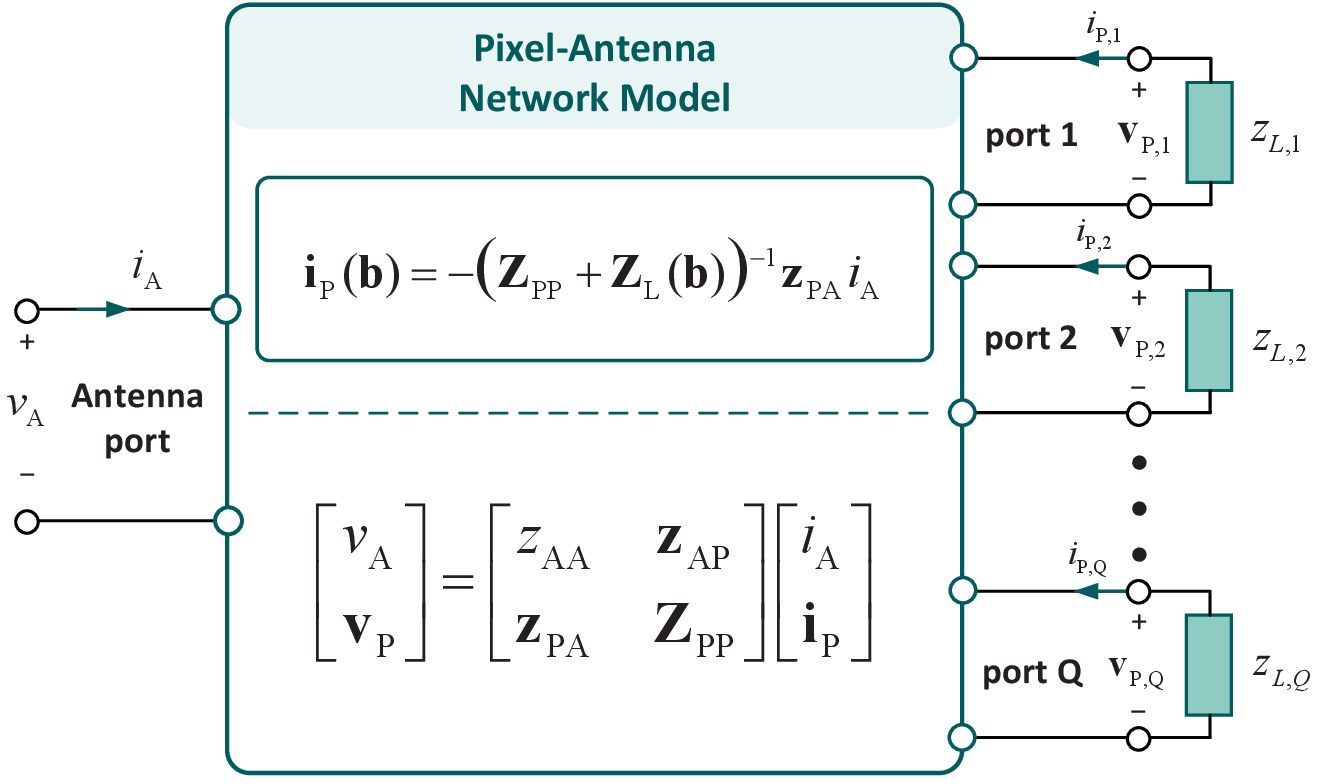}
  \caption{Model of the multiport circuit network.}
  \vspace{-0.0 cm}
  \label{fig:multiport_network}
\end{figure}

As depicted in Fig.~\ref{fig:multiport_network}, each pixel antenna consists of a radiating aperture discretized into parasitic metallic pixels. The interconnections between adjacent pixels are governed by electronic switches, allowing the antenna topology to be dynamically reconfigured through different switch states. Such topology reconfiguration alters the EM interactions among the pixels, thereby changing the induced current distribution and enabling manipulation of the radiated EM fields. Consequently, the radiated EM fields can be shaped to produce distinct radiation patterns and polarization states. 

To facilitate the analysis of the antenna radiation characteristics, we apply multiport circuit network theory to establish a unified modeling framework for the pixel antenna~\cite{Shen2025AntennaCodingPixel}. Specifically, the pixel antenna is modeled as a $(Q+1)$-port circuit network consisting of one antenna port for signal feeding and $Q$ pixel ports for inter-pixel connections. The multiport network is described by its impedance matrix, which is partitioned according to the antenna and pixel ports as
\begin{equation}
\mathbf Z \triangleq
\begin{bmatrix}
z_{\mathrm{AA}} & \mathbf z_{\mathrm{AP}}^{T}\\
\mathbf z_{\mathrm{PA}} & \mathbf Z_{\mathrm{PP}}
\end{bmatrix},
\label{eq:pixel_Z}
\end{equation}
where $z_{\mathrm{AA}}\in\mathbb C$ denotes the self-impedance of the antenna port, $\mathbf Z_{\mathrm{PP}}\in\mathbb C^{Q\times Q}$ denotes the mutual impedance matrix among the $Q$ pixel ports, and $\mathbf z_{\mathrm{AP}}\in\mathbb C^{Q}$ and $\mathbf z_{\mathrm{PA}}\in\mathbb C^{Q}$ characterize the trans-impedance between the antenna port and the pixel ports. For a reciprocal network, the off-diagonal impedance blocks satisfy $\mathbf z_{\mathrm{PA}}=\mathbf z_{\mathrm{AP}}$. Based on this impedance matrix, the voltage-current relationship among the ports can be expressed as
\begin{equation}
\begin{bmatrix}
v_{\mathrm A}\\
\mathbf v_{\mathrm P}
\end{bmatrix}
=
\begin{bmatrix}
z_{\mathrm{AA}} & \mathbf z_{\mathrm{AP}}^{T}\\
\mathbf z_{\mathrm{PA}} & \mathbf Z_{\mathrm{PP}}
\end{bmatrix}
\begin{bmatrix}
i_{\mathrm A}\\
\mathbf i_{\mathrm P}
\end{bmatrix},
\label{eq:pixel_VI}
\end{equation}
where $v_{\mathrm A}\in\mathbb C$ and $i_{\mathrm A}\in\mathbb C$ denote the voltage and current at the antenna port, respectively, and $\mathbf v_{\mathrm P}\in\mathbb C^{Q}$ and $\mathbf i_{\mathrm P}\in\mathbb C^{Q}$ denote the voltages and currents at the pixel ports. 
 
To describe the reconfigurable antenna state in a compact form, we introduce the binary vector $\mathbf b\triangleq[b_1,\ldots,b_Q]^T\in\{0,1\}^{Q}$ to specify the states of the switchable pixel ports. Each switch state determines the corresponding  load impedance of the  pixel port. For the $q$-th pixel port, the load impedance is given by
\begin{equation}
z_{\mathrm L,q}=
\begin{cases}
Z_{\mathrm{on}}, & b_q=0,\\
Z_{\mathrm{off}}, & b_q=1,
\end{cases}
\qquad q=1,2,\ldots,Q,
\label{eq:pixel_load}
\end{equation}
where \(b_q=0\) and \(b_q=1\) correspond to the ON/shorted and OFF/open states, respectively, and \(Z_{\rm on}\) and \(Z_{\rm off}\) denote the corresponding load impedances.

The diagonal load matrix associated with switch state vector $\mathbf b$ is written as $\mathbf Z_{\mathrm L}(\mathbf b)\triangleq
\mathrm{diag}(z_{\mathrm L,1},\ldots,z_{\mathrm L,Q})$.  Using the load termination condition at the pixel ports, the following relationship can be obtained
\begin{equation}\label{eq:load condition}
\mathbf v_{\mathrm P}=-\mathbf Z_{\mathrm L}(\mathbf b)\mathbf i_{\mathrm P}.
\end{equation}
Combining the multiport network equation in \eqref{eq:pixel_VI} with the load-termination condition in \eqref{eq:load condition}, we can derive the current vector at the pixel ports as
\begin{equation}
\mathbf i_{\mathrm P}(\mathbf b)=
-\big(\mathbf Z_{\mathrm{PP}}+\mathbf Z_{\mathrm L}(\mathbf b)\big)^{-1}\mathbf z_{\mathrm{PA}}\, i_{\mathrm A}.
\label{eq:pixel_i_collect}
\end{equation}
Accordingly, the currents at the antenna port and pixel ports can be stacked into the overall port-current vector $\mathbf i(\mathbf b)\triangleq[i_{\mathrm A}, \mathbf i_{\mathrm P}(\mathbf b)^{T}]^{T}\in\mathbb C^{Q+1}$.
\begin{figure}[!t]
  \centering
  \includegraphics[width=\linewidth]{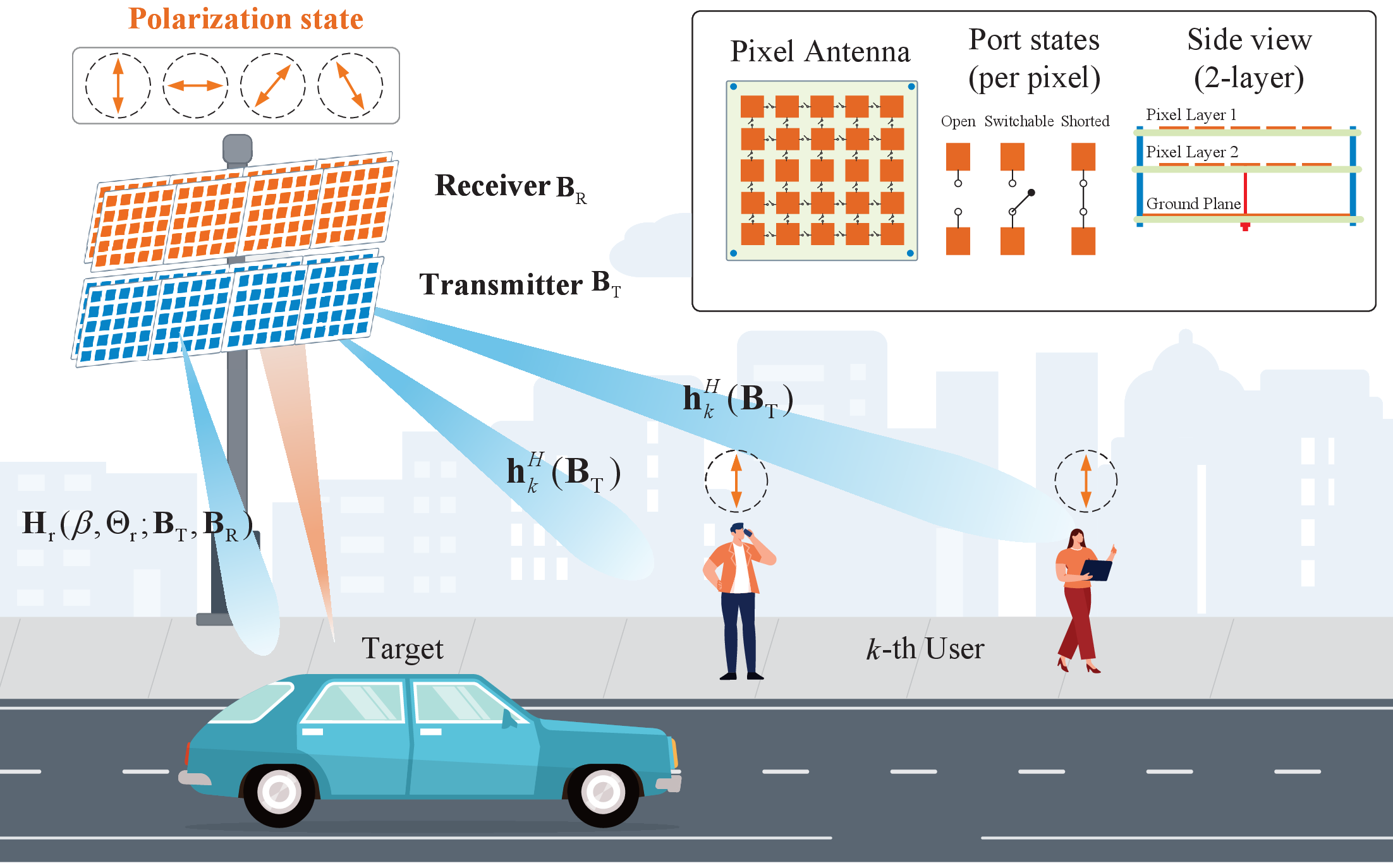}
  \caption{ISAC enabled by reconfigurable pixel-antenna arrays.}
  \label{fig:systemmodel}
\end{figure}
Given the port currents derived above, the overall radiation pattern can be formulated as a weighted sum of the open-circuit radiation patterns of all ports, i.e.,
\begin{equation}
\mathbf e(\mathbf b)
=
\mathbf e_{\mathrm A}i_{\mathrm A}
+
\sum_{q=1}^{Q}\mathbf e_{\mathrm P,q}\,i_{\mathrm P,q}(\mathbf b)
\triangleq
\mathbf E_{\mathrm{oc}}\,\mathbf i(\mathbf b),
\label{eq:pixel_e_superposition}
\end{equation}
where 
$\mathbf e_{\mathrm A}
\triangleq[\mathbf e_{\mathrm A,\theta}^{T},\mathbf e_{\mathrm A,\phi}^{T}]^{T}
\in\mathbb C^{2M}$ 
and 
$\mathbf e_{\mathrm P,q}
\triangleq[\mathbf e_{\mathrm P,q,\theta}^{T},\mathbf e_{\mathrm P,q,\phi}^{T}]^{T}
\in\mathbb C^{2M}$ 
are the open-circuit radiation patterns of the antenna port and the $q$-th pixel port, respectively. Each pattern vector contains the $\theta$- and $\phi$-polarized components over the $M$ sampled angular grid. These open-circuit patterns are obtained by exciting the corresponding port with a unit current while keeping all the remaining ports open-circuited. To ensure that different switch configurations are compared under the same radiation-energy level, the radiation response vector $\mathbf e(\mathbf b)$ is normalized to satisfy $\|\mathbf e(\mathbf b)\|=1$.  Moreover, the open-circuit radiation-pattern vectors of all ports can be collected into the following matrix
$\mathbf E_{\mathrm{oc}}
\triangleq[\mathbf e_{\mathrm A},\mathbf e_{\mathrm P,1},\ldots,\mathbf e_{\mathrm P,Q}]
\in\mathbb C^{2M\times(Q+1)}$.

\subsection{Pixel Antenna Array Modeling}
As shown in Fig.~\ref{fig:systemmodel}, we consider an ISAC scenario with $K$ single-antenna users and one target. The base station (BS) is equipped with separate transmit and receive arrays, both consisting of $N$ pixel antennas arranged as a uniform linear array (ULA) with adjacent-element spacing $d$. In contrast, each user is assumed to employ a conventional omnidirectional polarized antenna.

For pixel antenna arrays, the radiation characteristics of each antenna can be individually controlled by adjusting its switch configuration. We denote the switch state matrices of the transmit and receive pixel antenna arrays as
\begin{align}
\mathbf B_{\mathrm T}
&\triangleq[\mathbf b_{\mathrm T,1},\ldots,\mathbf b_{\mathrm T,N}]
\in\{0,1\}^{Q\times N},
\label{eq:BT_def}\\
\mathbf B_{\mathrm R}
&\triangleq[\mathbf b_{\mathrm R,1},\ldots,\mathbf b_{\mathrm R,N}]
\in\{0,1\}^{Q\times N},
\label{eq:BR_def}
\end{align}
where the binary vectors $\mathbf b_{\mathrm T,n}$ and $\mathbf b_{\mathrm R,n}$ denote the switch states associated with the $n$-th transmit and receive pixel antennas, respectively. Accordingly, the radiation pattern matrices of the transmit and receive pixel antenna arrays are
\begin{equation}
\mathbf E_{\mathrm T}(\mathbf B_{\mathrm T})
\triangleq
[\mathbf e(\mathbf b_{\mathrm T,1}),\ldots,\mathbf e(\mathbf b_{\mathrm T,N})]
\in\mathbb C^{2M\times N},
\label{eq:ET_un}
\end{equation}
\begin{equation}
\mathbf E_{\mathrm R}(\mathbf B_{\mathrm R})
\triangleq
[\mathbf e(\mathbf b_{\mathrm R,1}),\ldots,\mathbf e(\mathbf b_{\mathrm R,N})]
\in\mathbb C^{2M\times N}.
\label{eq:ER_un}
\end{equation} 
The array phase-response matrix over the $M$ sampled spatial angles is defined as $\mathbf A
\triangleq
[
\mathbf a(\theta_1),
\mathbf a(\theta_2),
\ldots,
\mathbf a(\theta_M)
]^T
\in\mathbb C^{M\times N}$, where the ULA steering vector is given by
$\mathbf a(\theta_m)
\triangleq
\frac{1}{\sqrt{N}}[
1,
e^{-\jmath{2\pi d\sin\theta_m}/{\lambda_{\mathrm c}}},
\ldots,
e^{-\jmath{2\pi (N-1)d\sin\theta_m}/{\lambda_{\mathrm c}}}]^T
\in\mathbb C^{N}$.
Since the same array phase response is shared by the
$\theta$- and $\phi$-polarized components, the stacked array
phase-response matrix is defined as $\bar{\mathbf A}
\triangleq [\mathbf A^T,\mathbf A^T]^T
\in\mathbb C^{2M\times N}$.
Accordingly, by jointly accounting for the radiation responses and the array phase differences, the effective transmit and receive array characteristic matrices can be written as
\begin{equation}
\mathbf G_{\mathrm T}(\mathbf B_{\mathrm T})
\triangleq
\mathbf E_{\mathrm T}(\mathbf B_{\mathrm T})\odot \mathbf{\bar{A}}
\in\mathbb C^{2M\times N},
\label{eq:GT_isac}
\end{equation}
\begin{equation}
\mathbf G_{\mathrm R}(\mathbf B_{\mathrm R})
\triangleq
\mathbf E_{\mathrm R}(\mathbf B_{\mathrm R})\odot \mathbf{\bar{A}}
\in\mathbb C^{2M\times N}.
\label{eq:GR_isac}
\end{equation}

\setcounter{TempEqnCnt}{\value{equation}}
\setcounter{equation}{20}

\begin{figure*}[b]
\hrulefill
\begin{equation}
\mathrm{SNR}_{\mathrm r}
=
{\operatorname{Tr}\!\Big\{
(\mathbf F^T\otimes \mathbf w^H)
\big(\mathbf G_{\mathrm T}^T(\mathbf B_{\mathrm T})
\otimes \mathbf G_{\mathrm R}^H(\mathbf B_{\mathrm R})\big)
\boldsymbol{\Omega}_{\mathrm V,\mathrm r}
\big(\mathbf G_{\mathrm T}^*(\mathbf B_{\mathrm T})
\otimes \mathbf G_{\mathrm R}(\mathbf B_{\mathrm R})\big)
(\mathbf F^*\otimes \mathbf w)
\Big\}}/{(\sigma_{\mathrm r}^2\mathbf{w}^{H}\mathbf w)}
.
\label{eq:SNRr_isac}
\end{equation}
\end{figure*}

\setcounter{equation}{\value{TempEqnCnt}}

Subsequently, we introduce the propagation channel model, which jointly characterizes the polarization-domain, the angular-domain, and the array phase responses of the wireless environment. To account for polarization rotation and mismatch in propagation, we adopt the reduced Jones-rotation model for both communication and sensing links:
\begin{equation}
\mathbf C_i(\Theta_i)
\triangleq
\begin{bmatrix}
\cos\Theta_i & -\sin\Theta_i\\
\sin\Theta_i & \cos\Theta_i
\end{bmatrix},
\label{eq:C_rot}
\end{equation}
where $i\in\{k,\mathrm{r}\}$, and $\Theta_i$ denotes the associated polarization rotation angles. The angular-domain channel is represented by $\mathbf H_{\mathrm A,i}\in\mathbb C^{M\times M}$, where each entry denotes the path gain from the discretized  angle of arrival (AoA) grid to the angle of departure (AoD). Combining this angular-domain channel with the polarization rotation model, the corresponding polarization-aware virtual channel matrix is modeled as
\begin{equation}
\mathbf H_{\mathrm V,i}
\triangleq
\mathbf C_i(\Theta_i)\otimes\mathbf H_{\mathrm A,i}
\in\mathbb C^{2M\times 2M}.
\label{eq:HVk_isac}
\end{equation} 
Based on the above definitions, the downlink channel from the BS to the $k$-th user is given by
\begin{equation}
\mathbf h_k(\mathbf B_{\mathrm T})
=\mathbf G_{\mathrm T}^{H}(\mathbf B_{\mathrm T})\mathbf H_{\mathrm V,k}^{H}
\mathbf e_{\mathrm{UE},k}
\in\mathbb C^{N},
\label{eq:hk_isac}
\end{equation}
where $\mathbf e_{\mathrm{UE},k}\in\mathbb C^{2M}$ denotes the radiation response vector of the receive antenna at the $k$-th user. 
For notational convenience, we define the Hermitian-form sensing matrix as
\begin{equation}
\mathbf H_{\mathrm r}(\mathbf B_{\mathrm T},\mathbf B_{\mathrm R})=\mathbf G_{\mathrm T}^{H}(\mathbf B_{\mathrm T})\mathbf H_{\mathrm V,\mathrm r}^{H}
\mathbf G_{\mathrm R}(\mathbf B_{\mathrm R})
\in\mathbb C^{N\times N}.
\label{eq:Hr_isac}
\end{equation}
Accordingly, the forward sensing channel from the transmit array to the receive array is represented by $\mathbf H_{\mathrm r}^{H}(\mathbf B_{\mathrm T},\mathbf B_{\mathrm R})$.

\subsection{Communication and Sensing Signal Models}
\label{subsec:signal_models}
With the channel model established, the received signal at the $k$-th user can be expressed as
\begin{equation}
y_k=\mathbf h_k^{H}(\mathbf B_{\mathrm T})\mathbf{F}\mathbf s+n_k,
\label{eq:yk_isac}
\end{equation}
where $\mathbf F
\triangleq
[\mathbf f_1,\ldots,\mathbf f_K,\mathbf f_{K+1},\ldots,\mathbf f_N]
\in\mathbb C^{N\times N}$ is the BB-domain beamformer with the total transmit power constraint $\mathrm{Tr}(\mathbf F\mathbf F^{H})\le P$, $\mathbf s\triangleq[\mathbf s_{\mathrm c}^{T}, \mathbf s_{\mathrm r}^{T}]^{T}\in\mathbb C^{N}$ denotes the symbol vector containing the communication symbols $\mathbf s_{\mathrm c}\in\mathbb C^{K}$ and radar probing signal $\mathbf s_{\mathrm r}\in\mathbb C^{(N-K)}$. These two signal vectors satisfy $\mathbb{E}\{\mathbf{s}_{\mathrm{c}}\mathbf{s}_{\mathrm{c}}^{H}\}=\mathbf{I}_{K}$ and $\mathbb{E}\{\mathbf{s}_{\mathrm{r}}\mathbf{s}_{\mathrm{r}}^{H}\}=\mathbf{I}_{N-K}$, respectively, and are assumed to be statistically independent. Here, $n_k \sim \mathcal{CN}(0,\sigma_k^2)$ denotes additive white Gaussian noise (AWGN) at the $k$-th user. 
The SINR of the $k$-th user  is given by
\begin{equation}
\mathrm{SINR}_k
=
\frac{|\mathbf h_k^{H}(\mathbf B_{\mathrm T})\mathbf f_k|^2}
{\sum_{\substack{j\neq k}}^{N}|\mathbf h_k^{H}(\mathbf B_{\mathrm T})\mathbf f_j|^2+\sigma_k^2}.
\label{eq:SINR_isac}
\end{equation}
For radar sensing, the BS applies receive combining to the echo signal, yielding the post-combining sensing signal as
\begin{equation}
y_{\mathrm r}
=
\mathbf{w}^{H}\mathbf H_{\mathrm r}^H(\mathbf B_{\mathrm T},\mathbf B_{\mathrm R})\mathbf{F}\mathbf s
+\mathbf w^{H}\mathbf n_{\mathrm r},
\label{eq:yr_isac}
\end{equation}
where $\mathbf w\in\mathbb C^{N}$ denotes the BB-domain combiner and $\mathbf n_{\mathrm r}\sim\mathcal{CN}(\mathbf 0,\sigma_{\mathrm r}^2\mathbf I_N)$ is AWGN at the BS. 
Therefore, the sensing SNR can be expressed as
\begin{equation}
\mathrm{SNR}_{\mathrm r}
\triangleq
\frac{
\mathbb E\big\{
\left|
\mathbf w^{H}\mathbf H_{\mathrm r}^{H}(\mathbf B_{\mathrm T},\mathbf B_{\mathrm R})
\mathbf F\mathbf s
\right|^2\big\}
}{
\mathbb E\big\{
\left|
\mathbf w^{H}\mathbf n_{\mathrm r}
\right|^2
\big\}
}.
\label{eq:sensing_snr_def_isac}
\end{equation}
\stepcounter{equation}
In many radar scenarios, while the target angles and ranges may remain static or slowly varying, the polarization scattering characteristics can fluctuate significantly with object orientation, surface properties, and small-scale scattering, and thus need to be modeled as random variables~\cite{liu2026clutter}. Therefore, we evaluate the sensing performance through the second-order statistics of the polarization-aware virtual sensing channel. Since the echo signal in~\eqref{eq:yr_isac} is propagated through $\mathbf H_{\mathrm r}^{H}(\mathbf B_{\mathrm T},\mathbf B_{\mathrm R})$, the statistical sensing SNR can be written as~\eqref{eq:SNRr_isac} shown at the bottom of this page.
Here, $\boldsymbol{\Omega}_{\mathrm V,\mathrm r}$ is the second-moment matrix of the virtual sensing channel vector $\mathbf h_{\mathrm V,\mathrm r}$, where $\boldsymbol{\Omega}_{\mathrm V,\mathrm r}
\triangleq
\mathbb E\left\{
\mathbf h_{\mathrm V,\mathrm r}
\mathbf h_{\mathrm V,\mathrm r}^{H}
\right\}$ and $\mathbf h_{\mathrm V,\mathrm r}\triangleq
\operatorname{vec}(\mathbf H_{\mathrm V,\mathrm r})$.


\subsection{Problem Formulation}

Based on the above model, we jointly optimize the EM-domain antenna switch configurations $\mathbf B_{\mathrm T}$ and $\mathbf B_{\mathrm R}$, the BB-domain beamformer $\mathbf F$ and combiner $\mathbf w$, to maximize the  sensing SNR while satisfying the communication QoS and power constraints. The optimization problem is formulated as
\begin{subequations}
\begin{align}
\max_{\mathbf B_{\mathrm T},\mathbf B_{\mathrm R},\mathbf F,\mathbf w}\quad
& \mathrm{SNR}_{\mathrm r}(\mathbf B_{\mathrm T},\mathbf B_{\mathrm R},\mathbf F,\mathbf w)\\
\mathrm{s.t.}\quad\quad
& \mathrm{SINR}_k(\mathbf B_{\mathrm T},\mathbf F)\ge \gamma_k,~ \forall k, \\
& \mathrm{Tr}(\mathbf F\mathbf F^{H})\le P,\\
& \|\mathbf w\|_2=1,\\
& \mathbf B_{\mathrm T},~\mathbf B_{\mathrm R}\in\{0,1\}^{Q\times N},
\end{align}
\label{eq:opt_isac_main}
\end{subequations}
where $\gamma_k$ denotes the required SINR of the $k$-th user.

\section{Joint EM-BB Beamforming Design}
In this section, we propose the joint EM-BB beamforming design that alternately optimizes the binary antenna switch states, the digital precoder, and the sensing combiner. Specifically, to effectively tackle the mixed-integer nonconvex problem \eqref{eq:opt_isac_main},  the binary antenna switch configurations  $\mathbf B_{\mathrm T}$ and $\mathbf B_{\mathrm R}$ are heuristically optimized using SEBO, the digital precoder $\mathbf{F}$ is updated via an MM-SOCP-based algorithm, and the sensing combiner is obtained in closed form by solving a Rayleigh quotient maximization problem.  The detailed algorithmic steps are presented as follows.

\subsection{EM-domain Beamforming Designs}
With the BB-domain beamforming matrices $\mathbf F$ and $\mathbf w$ fixed, the optimization problem for the EM-domain antenna switch states can be reformulated as
\begin{subequations}
\begin{align}
\max_{\mathbf B_{\mathrm T},\mathbf B_{\mathrm R}}\quad
& \mathrm{SNR}_{\mathrm r}(\mathbf B_{\mathrm T},\mathbf B_{\mathrm R})\\
\mathrm{s.t.}\quad\ 
& \mathrm{SINR}_k(\mathbf B_{\mathrm T})\ge \gamma_k, ~\forall k,\\
& \mathbf B_{\mathrm T},~\mathbf B_{\mathrm R} \in\{0,1\}^{Q\times N}.
\end{align}
\end{subequations}
To evaluate candidate antenna switch configurations in the ISAC system, we directly adopt a communication-aware selection rule, where only the candidates satisfying the communication QoS requirements are considered, and the one yielding the largest sensing SNR is selected. Accordingly, the corresponding evaluation function can be expressed as
\begin{equation}
\!\!\!\Phi(\mathbf B_{\mathrm T},\!\mathbf B_{\mathrm R})
\!=\!
\left\{
\begin{array}{@{}l@{\;}l@{}}
\mathrm{SNR}_{\mathrm r}(\mathbf B_{\mathrm T},\!\mathbf B_{\mathrm R}),&
\mathrm{SINR}_k(\mathbf B_{\mathrm T})\!\ge\!\gamma_k,\ \forall k,\\
-\infty,&
\mathrm{otherwise}.
\end{array}
\right.
\label{eq:direct_phi_hard_isac}
\end{equation}
However, this hard rule is overly conservative in the alternating optimization process. In particular, a trial EM-domain antenna switch configuration that violates the SINR constraints under the current BB-domain digital precoder $\mathbf F$ may still become feasible after the re-optimization. Therefore, directly discarding all currently infeasible trial switch configurations may prematurely exclude useful EM-domain antenna configurations. 
To avoid this issue, we use a refined communication-aware penalty merit function to rank binary switch configurations, which yields
\begin{equation}
\begin{aligned}
\!\!\!\!\!\Phi(\mathbf B_{\mathrm T},\!\mathbf B_{\mathrm R})
\!\triangleq\!
\mathrm{SNR}_{\mathrm r}(\mathbf B_{\mathrm T},\!\mathbf B_{\mathrm R})\!-\!\lambda\!\sum_{k=1}^{K}
\!\left[
\gamma_k\!-\!\mathrm{SINR}_k(\mathbf B_{\mathrm T})
\right]_+\!\!,\!\!\!
\end{aligned}
\label{eq:direct_phi_e_isac}
\end{equation}
where $[x]_+=\max(x,0)$, and $\lambda>0$ is a penalty factor for communication-QoS violation. Compared with the hard rule in \eqref{eq:direct_phi_hard_isac}, the formulation in \eqref{eq:direct_phi_e_isac} provides a softer evaluation criterion by penalizing the degree of communication-constraint violation rather than directly discarding infeasible candidates.

Since a full search over all EM-domain switch configurations requires $2^{2QN}$ trials, exhaustive enumeration is computationally prohibitive. Therefore, we adopt a SEBO strategy, where each column of the  binary antenna configurations $\mathbf B_{\mathrm T}$ and $\mathbf B_{\mathrm R}$ is partitioned into $G$ disjoint blocks and cyclically optimized via block-wise exhaustive search, with the remaining ones fixed~\cite{Shen2017SEBOPixel}. Let $Q_{\mathrm b} = \left\lceil {Q}/{G} \right\rceil$
denote the nominal block length. Then, the indices of the  switch states within the $g$-th block are included in  the set $\mathcal{J}_g$, which yields 
\begin{equation}
\mathcal J_g
\triangleq{}
\left\{
(g-1)Q_{\mathrm b}+1,\ldots,
\min(gQ_{\mathrm b},Q)
\right\}.
\label{eq:direct_block_partition_isac}
\end{equation}
The complementary set of $\mathcal J_g$  is given by
\begin{equation}
\overline{\mathcal J}_g
\triangleq{}
\{1,\ldots,Q\}\setminus\mathcal J_g.
\end{equation}
Starting from randomly initialized $\mathbf B_{\mathrm T}$ and $\mathbf B_{\mathrm R}$, we sequentially  optimize the transmit- and receive-side switch states by updating one antenna at a time, with the switch states of each selected antenna optimized block by block. Given the fixed switch states of all the other blocks, the $g$-th block in the $n$-th transmit antenna $\mathbf b_{\mathrm T,n,\mathcal J_g}$ is updated by 
\begin{equation}
\mathbf b_{\mathrm T,n,\mathcal J_g}
=
\arg\max_{\mathbf c\in\{0,1\}^{|\mathcal J_g|}}
\Phi
\left(
\mathbf c ;
\mathbf b_{\mathrm T,n,\overline{\mathcal J}_g},
\overline{\mathbf B}_{\mathrm T,n},
\mathbf B_{\mathrm R}
\right),
\label{eq:direct_tx_sebo_search_isac}
\end{equation}
where $\mathbf{c}\in\{0,1\}^{|\mathcal J_g|}$ represents the switch states to be optimized within the $g$-th block, $\mathbf b_{\mathrm T,n,\overline{\mathcal J}_g}$ denotes the fixed switch states of the remaining blocks in the $n$-th transmit antenna, and $\overline{\mathbf B}_{\mathrm T,n}$ denotes the switch states of all transmit antennas except the $n$-th one. After completing the update of the transmit-side antenna switch configurations $\mathbf B_{\mathrm T}$, the receive-side antenna switch configurations $\mathbf B_{\mathrm R}$ are updated in the same manner:
\begin{equation}
\mathbf b_{\mathrm R,n,\mathcal J_g}
=
\arg\max_{\mathbf c\in\{0,1\}^{|\mathcal J_g|}}
\Phi
\left(
\mathbf c ;
\mathbf b_{\mathrm R,n,\overline{\mathcal J}_g},
\overline{\mathbf B}_{\mathrm R,n},
\mathbf B_{\mathrm T}
\right).
\label{eq:direct_rx_sebo_search_isac}
\end{equation}

Compared with global exhaustive search, which requires
$2^{2QN}$ candidate evaluations, one SEBO sweep requires at most
$2NG2^{Q_{\rm b}}$ candidate evaluations, since the $G$ blocks of
each of the $N$ transmit and receive antennas are searched separately.
Thus, the enumeration burden of SEBO grows exponentially with the
block size $Q_{\rm b}$ rather than with the total number of switch
variables.
\vspace{-0.04in}
\subsection{BB-domain Beamforming Designs}
After designing the  EM-domain beamforming, we shift our focus to the design of the BB-domain beamforming. The optimization problem is formulated as   
\begin{subequations}
\begin{align}
\label{optimize_F_a}
\max_{\mathbf F,\mathbf w}\quad
& \mathrm{SNR}_{\mathrm r}(\mathbf F,\mathbf w)\\
\mathrm{s.t.}\quad
\label{problem_b}
& \mathrm{SINR}_k(\mathbf F)\ge \gamma_k, \forall k,\\
& \mathrm{Tr}(\mathbf F\mathbf F^{H})\le P,\\
& \|\mathbf w\|_2=1.
\end{align}
\end{subequations}
Based on the EM-domain antenna switch configurations, the radiation characteristics of the pixel antenna arrays can be determined. Accordingly, the effective channel of the target incorporating the radiation responses is defined as
\begin{equation}
\begin{aligned}
\!\!\!\boldsymbol{\Psi}
\!\triangleq\!
\left(
\mathbf G_{\mathrm T}^{T}(\mathbf B_{\mathrm T})
\!\otimes\!
\mathbf G_{\mathrm R}^{H}(\mathbf B_{\mathrm R})
\right)
\!\boldsymbol{\Omega}_{\mathrm V,\mathrm r}\!
\left(
\mathbf G_{\mathrm T}^{*}(\mathbf B_{\mathrm T})
\!\otimes\!
\mathbf G_{\mathrm R}(\mathbf B_{\mathrm R})
\right).\!
\end{aligned}
\label{eq:direct_Psi_isac}
\end{equation}
Thus, the objective $\mathrm{SNR}_{\mathrm{r}}$ in \eqref{optimize_F_a} can be further written as 
\begin{equation}
\mathrm{SNR}_{\mathrm r}
=
\frac{\operatorname{Tr}\!\left\{
(\mathbf F^{T}\otimes \mathbf w^{H})
\boldsymbol{\Psi}
(\mathbf F^{*}\otimes \mathbf w)
\right\}}{\sigma_{\mathrm r}^2\mathbf{w}^{H}\mathbf w}.
\end{equation}

\subsubsection{Digital Precoder Design}
The precoder \(\mathbf F\) is initialized by RZF or, if necessary, a feasibility SOCP. To facilitate the subsequent optimization of $\mathbf{F}$, we rewrite the objective function into the following equivalent form 
\begin{equation}\label{eq:SCNR_W}
\mathrm{SNR}_{\mathrm{r}}=\overline{\mathbf{f}}^{H}\overline{\mathbf{C}}_{\mathrm{F}}\overline{\mathbf{f}},
\end{equation}
where
\begin{equation}
\begin{aligned}
  \overline{\mathbf{f}} &\triangleq \mathrm{vec}(\mathbf{F}^{H}), 
  \\
  \overline{\mathbf{C}}_{\mathrm{F}}&\triangleq{\mathbf{C}_{\mathrm{F}}^{H}(\boldsymbol{\Psi}\otimes\mathbf{I}_{N})\mathbf{C}_{\mathrm{F}}}/
  {(\sigma_{\mathrm r}^2\mathbf{w}^{H}\mathbf w)},\\
  \mathbf{C}_{\mathrm{F}} &\triangleq (\mathbf{I}_{N}\otimes\mathbf{K}_{N,N})(\mathbf{I}_{N^{2}}\otimes\mathbf{w}). 
\end{aligned}
\end{equation}
Since $\overline{\mathbf{C}}_{\mathrm{F}}\succeq\mathbf 0$, the function $\overline{\mathbf{f}}^{H}\overline{\mathbf{C}}_{\mathrm{F}}\overline{\mathbf{f}}$ is convex in $\mathbf F$ , making the maximization problem nonconcave. Thus,  we utilize the MM method to relax it into a linear surrogate function
\begin{equation}
\begin{aligned}
\overline{\mathbf{f}}^{H}\overline{\mathbf{C}}_{\mathrm{F}}\overline{\mathbf{f}}\geq (\overline{\mathbf{f}}^{(t)})^{H}\overline{\mathbf{C}}_{\mathrm{F}}\overline{\mathbf{f}}^{(t)}+2\Re\{(\overline{\mathbf{f}}^{(t)})^{H}\overline{\mathbf{C}}_{\mathrm{F}}(\overline{\mathbf{f}}-\overline{\mathbf{f}}^{(t)})\}.
\end{aligned}
\label{eq:direct_sca_lb_F_isac}
\end{equation}
For the fractional communication SINR constraint in \eqref{problem_b}, we adopt the SOCP transformation, where the original SINR constraint is converted into a tractable second-order cone constraint with the desired signal constrained to be real-valued. Finally, the convex subproblem for optimizing the digital precoder $\mathbf{F}$ is given by 
\vspace{-0.1cm}
\begin{subequations}
\begin{align}
\!\!\!\!\!\!\max_{\mathbf F}\ 
& (\overline{\mathbf{f}}^{(t)})^{H}\overline{\mathbf{C}}_{\mathrm{F}}\overline{\mathbf{f}}^{(t)}+2\Re\{(\overline{\mathbf{f}}^{(t)})^{H}\overline{\mathbf{C}}_{\mathrm{F}}(\overline{\mathbf{f}}-\overline{\mathbf{f}}^{(t)})\}\\
\!\!\!\!\mathrm{s.t.}\ 
& \Re\{\mathbf h_k^{H}(\mathbf B_{\mathrm T})\mathbf f_k\}\!-\!\!\sqrt{\gamma_k}(\sum_{\substack{j\neq k}}^{N}|\mathbf h_k^{H}(\mathbf B_{\mathrm T})\mathbf f_j|^2\!\!+\!\sigma_k^2)^{\frac{1}{2}}
\!\!\geq\!0,\!\!\!\!\!\!\\
& \Im\{\mathbf h_k^{H}(\mathbf B_{\mathrm T})\mathbf f_k\}=0,\\
& \operatorname{Tr}(\mathbf F\mathbf F^{H})\le P,
\end{align}
\label{eq:direct_socp_F_isac}
\end{subequations}
which can be efficiently tackled by CVX.

\subsubsection{Digital Combiner Design}
Similarly, to facilitate the optimization of the digital combiner $\mathbf w$, we recast the objective function into an equivalent tractable form, leading to the following simplified subproblem 
\begin{subequations}
\label{eq:original_problem_u}
\begin{align}
    \max_{\mathbf{w}}&\ \ \mathbf{w}^{H}\overline{\mathbf{C}}_{\mathrm{w}}\mathbf{w},\\
    \mathrm{s.t.}& \ \ \|\mathbf w\|_2=1,
\end{align}
\end{subequations}
where we define 
\begin{equation}
\begin{aligned}
  \overline{\mathbf{C}}_{\mathrm{w}} &\triangleq \frac{1}{\sigma_{\mathrm r}^2}\mathbf{C}_{\mathrm{w}}^{H}(\boldsymbol{\Psi}\otimes\mathbf{I}_{N})\mathbf{C}_{\mathrm{w}},\\
   \mathbf{C}_{\mathrm{w}} &\triangleq (\mathbf{I}_{N}\otimes\mathbf{K}_{N,N})(\mathrm{vec}(\mathbf{F}^{H})\otimes\mathbf{I}_{N}).
\end{aligned}
\end{equation}
It can be observed that problem \eqref{eq:original_problem_u} takes the form of a Rayleigh quotient maximization problem, whose optimal solution is given by the eigenvector associated with the largest eigenvalue of $\overline{\mathbf{C}}_{\mathrm{w}}$, which is then normalized to satisfy the unit-norm constraint:
\vspace{-0.1cm}
\begin{equation}
\begin{alignedat}{2}
&\overline{\mathbf w} &&=
\mathrm{eigvec}_{\max}\!\left(\overline{\mathbf C}_{\mathrm w}\right),\\[-2pt]
&\mathbf w^\star &&=
\frac{\overline{\mathbf w}}{\|\overline{\mathbf w}\|_2}.
\end{alignedat}
\label{eq:direct_w_update_isac}
\end{equation}


\begin{algorithm}[!t]

\caption{Joint EM-BB Beamforming Design}
\label{alg:direct_sebo_ao_isac}
\begin{algorithmic}[1]
\STATE \textbf{Input:} $N$, $Q$, $K$, $\mathbf Z$, $\mathbf E_{\mathrm{oc}}$, $Z_{\mathrm{on}}$, $Z_{\mathrm{off}}$, $P$, $\lambda$, $G$, $\boldsymbol{\Omega}_{\mathrm V,\mathrm r}$, $\{\mathbf H_{\mathrm V,k}\}_{k=1}^{K}$, $\sigma_{\mathrm r}^2$, $\{\sigma_k^2\}_{k=1}^{K}, \{\gamma_k\}_{k=1}^{K}$.
\STATE \textbf{Output:} $\mathbf B_{\mathrm T}$, $\mathbf B_{\mathrm R}$, $\mathbf F$, $\mathbf w$.
\STATE Initialize $\mathbf B_{\mathrm T}$, $\mathbf B_{\mathrm R}$, $\mathbf F$.
\STATE Compute $\boldsymbol{\Psi}$ by~\eqref{eq:direct_Psi_isac} and update the combiner $\mathbf w$ by~\eqref{eq:direct_w_update_isac}.
\WHILE{no convergence of the sensing SNR}
    \STATE Sequentially update the EM-domain switch configurations $\mathbf B_{\mathrm T}$ and $\mathbf B_{\mathrm R}$ by SEBO according to~\eqref{eq:direct_tx_sebo_search_isac} and~\eqref{eq:direct_rx_sebo_search_isac}.
    \STATE Update $\boldsymbol{\Psi}(\mathbf B_{\mathrm T},\mathbf B_{\mathrm R})$ by~\eqref{eq:direct_Psi_isac}.
    \STATE Update the BB-domain beamformer $\mathbf F$ by~\eqref{eq:direct_socp_F_isac}.
    \STATE Update the combiner $\mathbf w$ by~\eqref{eq:direct_w_update_isac}.
\ENDWHILE
\STATE \textbf{Return:} $\mathbf B_{\mathrm T}$, $\mathbf B_{\mathrm R}$, $\mathbf F$, and $\mathbf w$.
\end{algorithmic}
 \vspace{0.2cm}   
\end{algorithm}

\section{Simulation Results}

We evaluate the proposed joint EM-BB beamforming design using CST-based antenna data and link-level simulations. The carrier frequency is $f_\mathrm{c}=2.45$ GHz. The BS employs separate transmit and receive arrays, each with $N=4$ antenna elements, and serves $K=2$ users. Unless otherwise specified, the user and target angles are drawn from $[-50^\circ,50^\circ]$, while the user and target distances are drawn from $[40,50]$ m and $[30,40]$ m, respectively. A line-of-sight propagation model is considered with distance-dependent power gain $\beta_i(d_i)=10^{-C_0/10}(d_i/D_0)^{-\kappa_i}$, where $C_0=30$ dB, $D_0=1$ m, and $\kappa_i=2.2$. The polarization rotation angles are drawn from $[-90^\circ,90^\circ]$. The SINR threshold is $\gamma_k=10$ dB for all users, and the user and radar noise powers correspond to $-70$ dBm.

\begin{figure}[!t]
    \centering
    \makebox[\linewidth][c]{%
        \includegraphics[width=3.2 in]{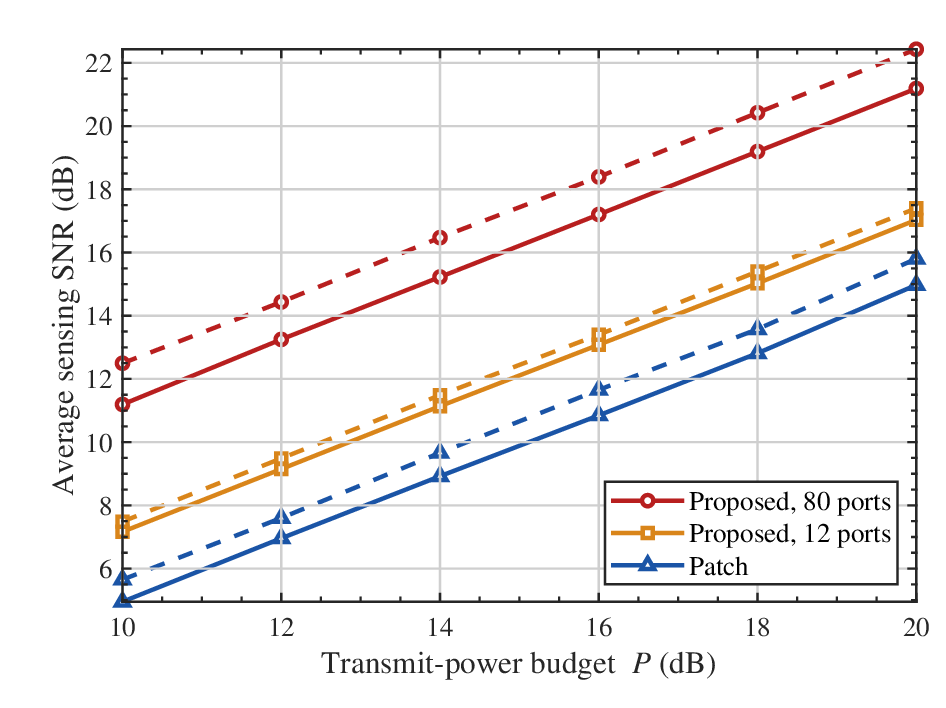}%
    }
    \caption{Average sensing SNR versus transmit-power budget \(P\) (Solid lines: ISAC; Dashed lines: Radar-only).}
    \label{fig:sensing_snr_vs_power}
\end{figure}

For the pixel antenna, the impedance matrix $\mathbf Z$ and the open-circuit radiation-pattern matrix $\mathbf E_{\rm oc}$ are extracted from CST full-wave simulations. The angular grid is sampled with a resolution of $5^\circ$ in the considered cutting plane, yielding $M=72$ angular samples. Based on the same $Q=80$ pixel-antenna model, we evaluate the proposed EM-BB beamforming design with two switch-configurable schemes: ``\textbf{Proposed, 80 ports}'', which optimizes all 80 controllable ports, and ``\textbf{Proposed, 12 ports}'', which optimizes 12 preselected ports following \cite{Zheng2025PolarizationPixel} with the remaining switch states fixed. This comparison illustrates the performance--complexity tradeoff of switch-state optimization. The SEBO block size is set to $Q_{\rm b}=5$. A conventional patch antenna array with the same aperture, optimized only in the BB domain, is used as the fixed-radiation benchmark and denoted as ``\textbf{Patch}''. For each architecture, a radar-only counterpart is included as a sensing-oriented baseline.
\begin{figure}[!t]
    \centering
    \makebox[\linewidth][c]{%
        \includegraphics[width=3.2 in]{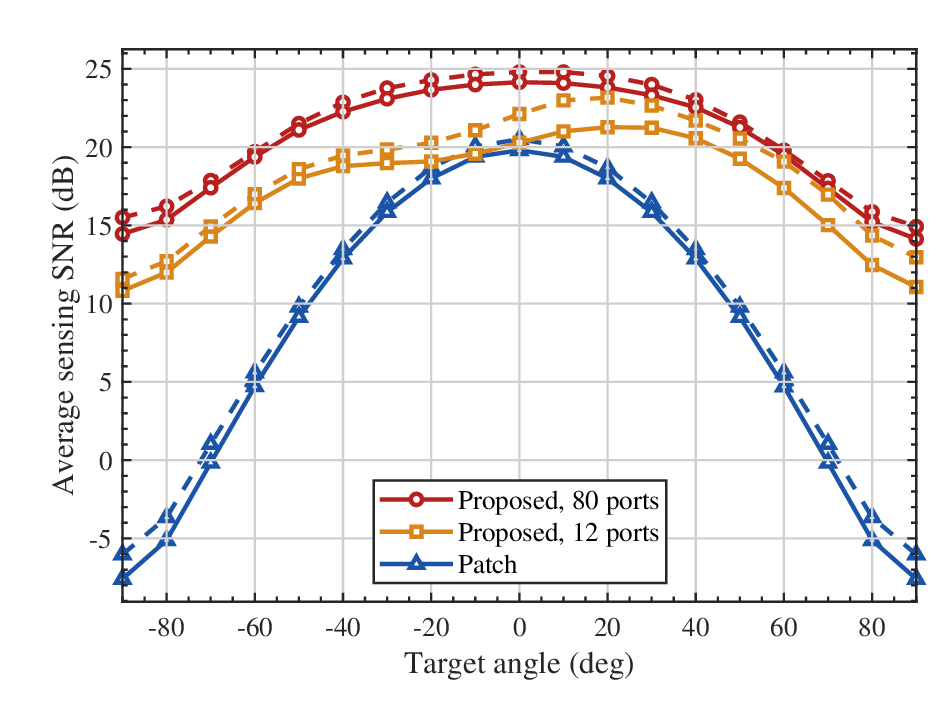}%
    }
    \caption{Average sensing SNR versus target angle (Solid lines: ISAC; Dashed lines: Radar-only).}
    \label{fig:angle versus SNR}
\end{figure}
Fig.~\ref{fig:sensing_snr_vs_power} shows the average sensing SNR versus transmit-power budget. The sensing SNR increases with power for all schemes due to stronger target illumination and echo reception. The proposed pixel-antenna schemes outperform the fixed-radiation patch benchmark in both ISAC and radar-only cases, showing the benefit of switch-configurable EM responses beyond BB-domain beamforming. The radar-only curves are higher without user-SINR constraints. Moreover, the 80-port scheme consistently outperforms the 12-port scheme, since more optimized switch states  enable finer pattern/polarization adaptation. This gain comes with higher hardware-control and optimization complexity, while the 12-port scheme provides a lower-complexity alternative with reduced but still evident sensing gains over the patch benchmark.

Fig.~\ref{fig:angle versus SNR} shows the average sensing SNR versus target angle. The fixed-radiation patch benchmark suffers severe SNR degradation at off-boresight angles, since its EM response cannot adapt to the target direction. In contrast, the proposed pixel-antenna schemes maintain much higher sensing SNR over a wide angular range by optimizing switch-configurable pattern and polarization responses. The 80-port scheme further improves the SNR stability over the 12-port scheme due to more controllable switch states. These results verify the target-angle robustness of the proposed joint EM-BB beamforming design.

\section{Conclusion}
This paper investigated joint EM-BB beamforming for ISAC systems with pattern- and polarization-reconfigurable pixel-antenna arrays. A multiport-network-based antenna model was integrated with polarization-aware communication and statistical sensing channels, yielding a mixed-integer nonconvex problem for average sensing-SNR maximization under user SINR and transmit-power constraints. To solve this problem, a SEBO-based alternating algorithm was developed. CST-based simulations showed that the proposed design improves sensing SNR and target-angle robustness over the fixed-patch benchmark, with larger gains achieved by increasing the number of controllable ports.

\end{document}